\documentclass[%
preprint,
 amsmath,amssymb,
 aps,
pra,
showkeys, superscriptaddress, onecolumn, longbibliography 
]{revtex4-2}

\usepackage{graphicx}
\usepackage{dcolumn}
\usepackage{bm}
\usepackage{hyperref}

\usepackage[symbol]{footmisc}

\usepackage{xcolor}

\begin{document}


\title{Quantitative coarse graining of laminar fluid flow penetration in rough boundaries}


\author{Akankshya Majhi}
\affiliation{Physical Chemistry and Soft Matter, \\ Wageningen University and Research, Wageningen, The Netherlands}
 
\author{Lars Kool}
\affiliation{Physical Chemistry and Soft Matter, \\ Wageningen University and Research, Wageningen, The Netherlands}
\affiliation{Laboratoire de Physique et Mécanique des Milieux Hétèrogènes,\\ École supérieure de physique et de chimie industrielles de la Ville de Paris,\\ Paris, France}


\author{Jasper van der Gucht}
\affiliation{Physical Chemistry and Soft Matter, \\ Wageningen University and Research, Wageningen, The Netherlands}

\author{Joshua A. Dijksman{\footnote{*Corresponding author: joshua.dijksman@wur.nl}}}
\affiliation{Physical Chemistry and Soft Matter, \\ Wageningen University and Research, Wageningen, The Netherlands}



 

\date{\today}

\begin{abstract}
The interaction between a fluid and a wall is described with a certain boundary condition for the fluid velocity at the wall. To understand how fluids behave near a rough wall, the fluid velocity at every point of the rough surface may be provided. This approach requires detailed knowledge of, and likely depends strongly on the roughness. Another approach of modeling the boundary conditions of a rough wall is to coarse grain and extract a penetration depth over which on average the fluid penetrates into the roughness. In this work we show that for a broad range of periodic roughness patterns and relative flow velocities, a universal penetration depth function can be obtained. We obtain these results with experiments and complementary numerical simulations. Our results show that wall roughness boundary conditions can be captured with an average ``slip length'' and so indicate that surface patterning yields extensive control over wall slip.
\end{abstract}

\maketitle


\section{\label{Intro}Introduction}

The interaction of a fluid with a wall can be very complex, especially when the wall possesses small-scale features \cite{Bottaro2019}. Typically, complexities arise in the form of slip layers \cite{Chen2012, Gao2018} and other complicated boundary conditions \cite{Achdou1998}. These wall-fluid interactions can, however, be tuned to control the fluid flow behaviour further away from the wall. The walls of the flow geometry can be modified by introducing roughness in the form of riblets or grooves on their surface. 

The concept of a slip length goes back to Navier, but the study on explicitly patterned ribbed surfaces came to prominence around 40 years ago, with inspiration from nature \cite{Bottaro2019, Bechert1989,Bechert2000, Bixler2013}. Sharks have denticles on their skin which reduces drag and allows for easy  swimming. To understand the effect of drag reduction in animals, work has been done on simplified model systems with various designs for ridges such as L-shaped, V-shaped and U-shaped \cite{Abdulbari2013, Raayai-Ardakani2019, Djenidi1989}. These studies primarily deal with variations in ridge dimensions such as ridge depth, ridge spacing, ridge width and angle of ridge orientation and in particular, their effect on the performance of the ridged surfaces \cite{Abdulbari2013, Raayai-Ardakani2019, Soleimani2021, Cafiero2022}. Since then, several studies have been carried out on engineered patterned ribbed geometries in parallel plate and concentric cylindrical systems. These patterned geometries have been studied in the context of reducing viscous (friction) drag \cite{Walsh1982,Walsh1983,Walsh1984, Walsh1990, Bechert1997, Davies2006, Gruneberger2011, Bixler2013, Raayai-Ardakani2017, Raayai-Ardakani2019, Raayai-Ardakani2020,Owens2020} or suppressing wall slip \cite{Nickerson2005, Chen2012, Luchini1991}. Walsh \cite{Walsh1983} indicated that microsurface geometry variations change the near-wall structure of the flow turbulent boundary layer and thus, are effective in reducing (viscous) drag. Bechert \textit{et al.} \cite{Bechert1997, Bechert1989} carried out an extensive parametric study on surfaces with longitudinal ribs, where they argued that the velocity profile in between the ridges penetrates to a distance below the ridge tips which they refer as ``protrusion height''. This protrusion height depended on the ridge dimensions and was indicative of the drag reduction. Despite showing interesting results, their geometries posed a problem in terms of manufacturing and durability.  Davies \textit{et al.} \cite{Davies2006} numerically studied the effect of patterned channel walls with alternating microribs and cavities. They investigated the influence of the vapour cavity depth in the entire laminar flow regime and showed significant reduction in the frictional resistance in laminar fluid flow. Their experimental studies were documented in \cite{Woolford2009}. Experimental work by Maynes \textit{et al.} \cite{Maynes2007} investigated the laminar flow in a parallel plate microchannel with ultrahydrophobic top and bottom walls and indicated dramatic decrease in the overall flow resistance. Djenidi \textit{et al.} \cite{Djenidi1989, Djenidi1994} showed that riblets cause a reduction in frictional drag in laminar flows. These drag reducing techniques find applications in aerospace industry for saving fuel costs \cite{Walsh1982, Walsh1983,Walsh1984, Walsh1990} and also serve as a non-additive drag reduction technology in pipe flow.

Even very recently, McKinley and co-workers \cite{Raayai-Ardakani2020} used different surface microtextures to study the drag reducing effect on the fluid flow over a range of flow speeds. They examined the interaction of Taylor vortices with the riblets in case of Newtonian fluids. Nickerson and Kornfield \cite{Nickerson2005} have shown that cleated surfaces on parallel plate geometry can be used to suppress wall slip. Although most of the previous researchers have established that rough surfaces can often be used to prevent wall slip, for instance in rheological measurements, so far they have not considered that this wall roughness also modifies the flow of the fluid near the wall, either by introduction of secondary flows or by penetration of the fluid flow into the gaps between the ridged wall. Additionally, riblets used are typically very complex in structure and the flow fields in such geometries are not easy to study systematically. Such complexities make interpretation of rheological measurements on the effect of rough walls difficult.

\begin{figure}[htbp!]
\centering
\includegraphics[width=1.0\linewidth]{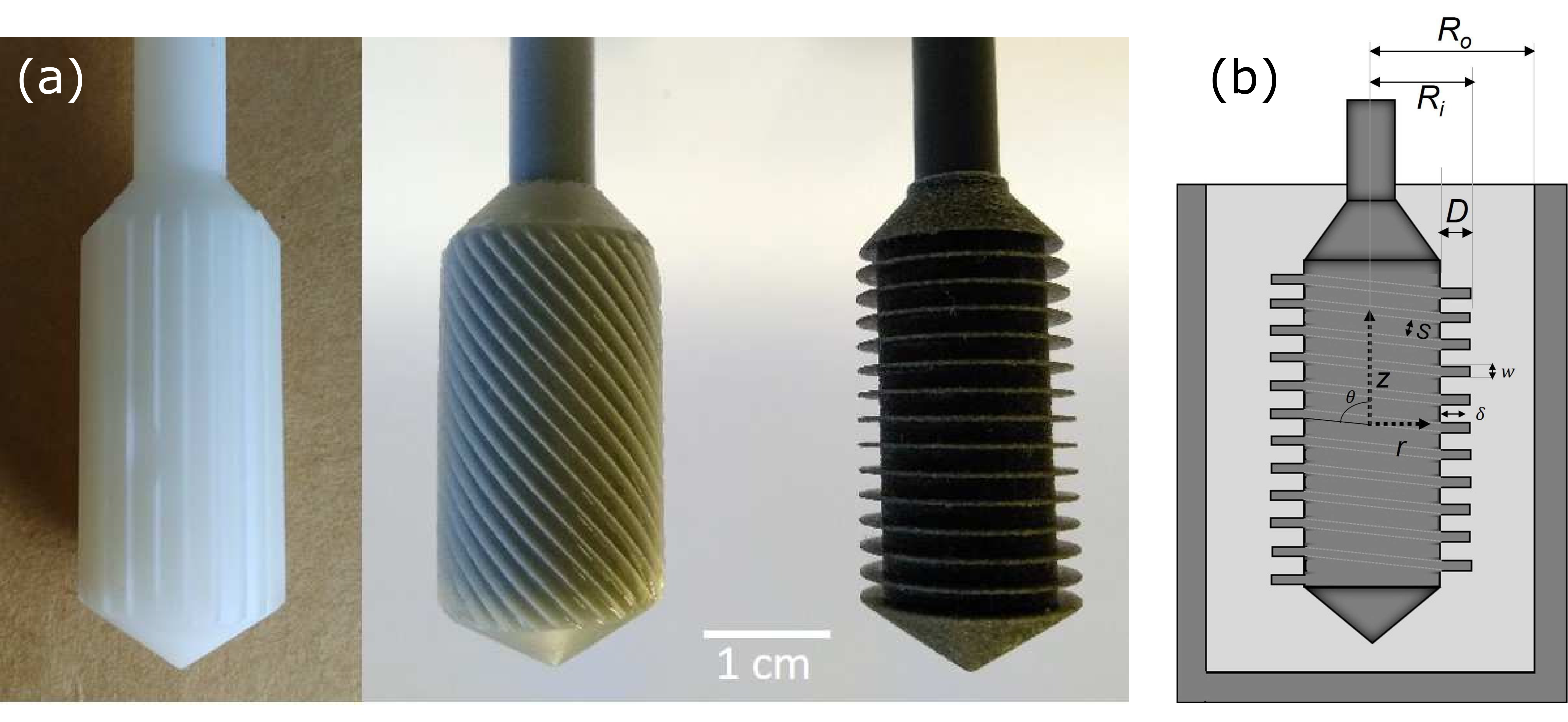}
\caption {(a) 3D printed geometries with patterned surfaces: vertical ridges (left), angled ridges (centre) and horizontal ridges (right). (b) Schematic representation of a concentric cylinder (CC) geometry with angled ridges. $R_i$ and $R_o$ are the radii of the inner and outer cylinder of the ridged geometry, respectively. Here the inner radius $R_i$ corresponds to the distance from the center of the coordinate system to the farthest point on the ridges. $D$ is the ridge depth, $S$ is spacing between the ridges, $w$ is the ridge width.}
\label{ridged_geo_actual&schematic}
\end{figure}

In the present paper, we show that the role of the wall roughness on the boundary layer of a laminar fluid flow can be effectively quantified by an effective penetration depth for a range of different roughness conditions. We use simple 3D printed ridged concentric cylinder geometries with ridges of different depth, spacing and orientation. This way of creating rough surfaces gives the ability to systematically investigate the role of a ridge in fluid flow, to demonstrate how the flow penetrates between the ridges and to examine whether there are effects of secondary flows, orthogonal to the primary flow direction. Fig.~\ref{ridged_geo_actual&schematic}(a) shows examples of patterned geometries. $D$, $S$  and $w$ are the ridge depth, ridge spacing, ridge width, respectively. The ridge angle $\theta$ can be varied with respect to the vertical axis. A helical wall patterning is shown to induce secondary flows at all flow rates, similar to Taylor vortices which are usually observed at high rotation rates as indicated by the Taylor number \cite{Faber1995}. 

In this work, we investigate the influence of 3D printed ridged geometries on the flow of a Newtonian fluid at low Reynolds number using standard concentric cylinder rheology. 
In Section \ref{MnM}, we first describe the materials, and experimental and numerical methods used for the determination of protrusion height. Later in Section \ref{RnD}, we discuss both numerical and experimental results of this flow profile penetration. We first present the simulated flow profiles in different ridged geometries. Then we explore the flow profile penetration, both experimentally and numerically, for two limiting cases of horizontal and vertical ridges and then discuss the angled ridge case.

\section{\label{MnM}Materials and Methods}

\subsection{Design and fabrication of ridged geometries}

Custom 3D printed ridged geometries were fabricated on a Stratasys Objet30 Scholar and a Formlabs 2 SLA (Formlabs, Inc.) printer to fit the Anton Paar rheometer shaft for disposable geometries.

This way of patterning with 3D printing allows to create well-defined geometries, compared to other manufacturing methods, with a good control to produce fine ($<$1 mm) features with a dimensional accuracy of $\sim$100 $\mu$m. The additional benefit of using the 3D printed ridged geometries is that they can be easily incorporated into a flow imaging technique such as MRI \cite{Callaghan1999,Coussot2020}, which allows to fully quantify the flow profiles and extract the wall stresses present in the system.

The geometries with ridges perpendicular to the rotational axis were designed by making cutouts of various spacings (0.3$\--$3 mm) and depths (0.5$\--$3 mm) from the standard CC17 geometry  with constant $R_i=8.5$ mm and $R_o=9$ mm. They were printed on the Objet30 Scholar, using the Vero Black photopolymer (Stratasys) with water-soluble supports. The geometries were printed vertically with a layer thickness of 16 $\mu$m.

The geometries with ridges at an angle with respect to the rotational axis were designed with the angle ranging from 10$^\circ$ to 60$^\circ$ with 10$^\circ$ intervals. Geometries with ridge angles between 60$^\circ$ to 90$^\circ$ could not be printed due to limitations of the printing technique. The geometries were designed such that the shortest distance (spacing) between the ridges was 2 mm, with a ridge depth of 1 mm and ridge thickness of 0.3 mm. This was achieved by changing the width and the number of cutouts. The width and the number of cutouts were calculated using a Python script with the length and the diameter of the ridged geometry, ridge angle and shortest distance between the ridges as input. These geometries with angled ridges were fabricated on a Formlabs 2 SLA printer, using the glass reinforced Rigid resin (Formlabs, Inc.) with a layer height of 50 $\mu$m and xy-resolution of 140 $\mu$m. The support needed for the printing was only attached to the shaft and the top chamber to prevent the introduction of artifacts coming from small residuals of the support material.

After printing, the geometries were washed twice for 10-15 minutes in 95$\%$ isopropanol. After drying, the geometries were post-cured under a 366 nm lamp for 1.5-2 hours. The geometries were rotated 180$^\circ$ halfway through to cure evenly. The support material was removed manually using flush side-clippers after curing. The shaft of all geometries was printed with a diameter of 8 mm and milled down on a lathe to 6.95 mm to fit the Anton Paar shaft for disposable geometries. We determined that this lathe post-processing ensured the best concentricity of the geometry with the Anton Paar rheometer shaft and cup.

\subsection{Rheological experiments}

Rheology experiments were performed on Anton Paar MCR 300 and MCR 501 rheometers, using the custom geometries in standard cups and cup holders (Anton Paar CC17). The experiments were performed at 20.5 $^\circ$C, using either a Peltier heat exchange element and waterbath at 20~$^\circ$C as heat sink (for MCR 501) or a high flowrate waterbath at 20.5 $^\circ$C (for MCR 300). All the printed geometries were tested on a Newtonian fluid. We used castor oil as a Newtonian fluid, due to its high viscosity ($\sim$1 Pa s at 20 $^\circ$C compared to 50 mPa s for most oils) and its stability over time. Unlike other viscous fluids like glycerol, castor oil is non-hygroscopic and it does not evaporate at room temperature, does not degrade over time, and does not swell nor degrade the 3D printed geometries (contrary to most organic solvents).

Before each measurement, the geometry was placed in a container with the test fluid to eliminate air bubbles between the ridges. The geometries were then visually inspected for the presence of air bubbles, which, if present, were removed by rubbing the geometry against the cup while it was submerged in the fluid.

We performed steady shear measurements with average shear rates ranging from $\dot{\gamma}$ = 10$^{-3}$  to 10$^{2}$ s$^{-1}$, resulting in a range of Re from 2.5 $\times$ 10$^{-5}$ to 2.5 for castor oil (only the laminar regime was probed). Each measurement consisted of 26 datapoints, distributed logarithmically, and each datapoint started with a stepwise increase of $\dot{\gamma}$, followed by an equilibration period of maximum 2 min, after which the data was averaged for 5 s. For each measurement, the samples were presheared at $\dot{\gamma}$ = 1000 s$^{-1}$ for 60 s followed by a recovery period of 60 s, to ensure that the stress from loading the sample in the measurement geometry did not influence the results and also to establish reproducible initial conditions.

To test the viability of using 3D printed geometries in rheology measurements, a solid CC geometry was designed with an outer diameter as close as possible to that of a commercially available smooth standard stainless steel CC geometry. These geometries were checked by comparing the flow curves of the test fluid. Both the 3D printed and the standard geometries yielded identical flow curves indicating that the accuracy of the 3D printer was high enough to produce complex rheology geometries and no deviations were expected in the results due to the intrinsic roughness of the 3D printed geometries. 
%
%
%
\subsection{Simulations}
\label{simulations}

Finite element simulations were done using COMSOL Multiphysics 5.6 to complement the experiments. To simplify the calculations, the ridged geometries were approximated as infinitely large parallel plate geometries. We performed simulations for varying $S$, $D$, and $\theta$, while keeping the gap size $R_\textnormal{gap,0}$ (i.e. the distance between the outer wall and the tip of the ridges) and ridge width $w$ constant and equal to the experimentally used values. We then impose a constant velocity difference $\Delta v$ between the inner and outer wall, and solve the Stokes equation (hence, ignoring inertial effects) numerically using quadratic quadrilateral elements for the velocity components and linear elements for the pressure. Since the flat plate limit does not account for the finite curvature of the actual CC geometry,  we performed additional simulations for curved geometries with horizontal and vertical ridges to confirm that the effects of curvature were small. 

\subsection{Determination of penetration depth}
\label{delta_calcmethod}

When a Newtonian fluid is sheared between two surfaces at a certain shear rate $\dot{\gamma}$, a shear stress $\sigma$ is generated and it can be defined as $\sigma = \eta\dot{\gamma}$, with $\eta$ being the viscosity of the fluid. The fluid flow behaviour can be modified using different boundary conditions, imposed by employing different geometries in a rheometer. Concentric cylinder (CC) is one of the most commonly used geometries, where the fluid is sheared in the gap between the inner and outer cylinder of radius $R_i$ and $R_o$, respectively. The inner cylinder is made to rotate at a rotational velocity $\Omega_i$ in revolutions per second (rps). The average shear rate $\dot{\gamma}$, in s$^{-1}$, is defined as the surface velocity of the moving inner cylinder divided by the gap size $R_\textnormal{gap}$ (Eq.~(\ref{shearrate_exp})). It should be noted that the shear rate is not constant within the gap.
\begin{equation}
\label{shearrate_exp}
\dot{\gamma} = \frac{2 \pi R_i \Omega_i}{R_\textnormal{gap}} \hspace{1cm} \text{where} \hspace{1cm} R_\textnormal{gap} = R_o - R_i
\end{equation}
The torque $M$ acting across the cylindrical surface is given by Eq.~(\ref{torqueeqn}), where $\omega_i$ is the angular velocity in rad s$^{-1}$ \citep{Landau1987, Macosko1994}. 
\begin{equation}
M = \frac{4 \pi \eta \omega_i L (R_o R_i)^2}{R_o ^2 - R_i ^2} \hspace{0.5cm} \text{where} \hspace{0.5cm} \omega_i = 2 \pi \Omega_i
\label{torqueeqn}
\end{equation}
The shear force in the azimuthal direction $F_\textnormal{shear}$ induced by the fluid can be expressed as the ratio of $M$ and the radius of the inner cylinder $R_i$. Thus, the shear stress in the azimuthal direction $\sigma$ can be expressed as $F_\textnormal{shear}$ divided by the surface area of the inner cylinder $A_i$ (Eq.~(\ref{shearstress_exp})). 
\begin{equation}
\label{shearstress_exp}
\sigma = \frac{F_\textnormal{shear}}{A_i}  =  \hspace{0.1cm} \frac{M}{2 \pi R_i^2 L} \hspace{0.1cm}  \text{where} \hspace{0.2cm} F_\textnormal{shear} = \frac{M}{R_i} \hspace{0.1cm} , \hspace{0.3cm} A_i = 2 \pi R_i L
\end{equation}

When roughness is introduced into the CC geometry in the form of ridges (as shown in Fig.~\ref{ridged_geo_actual&schematic}(b)), the flow profiles penetrate in the space between the ridges. We now make the step to quantify this flow profile penetration not via the flow field, but via the effect it has on the effective stress on the wall. We denote the extent of this penetration by a parameter called the penetration depth $\delta$, similar to the protrusion height as given in \cite{Bechert1989, Dean2010, Bechert1997, Luchini1991, Lee2001}. Typically, the penetration depth is determined by inducing fluid flow over a flat surface with ridges on it. In their theoretical investigation, Bechert and coworkers \cite{Bechert1989} simulated velocity distributions for various  ridge configurations and found that the apparent origin of the velocity profile in between the ridges lies in the gap between the inner flat surface and the tip of the ridge. Lee and Lee \cite{Lee2001} experimentally investigated the flow structures inside the semi-circular ridges. They referred to the distance between this origin and the ridge tip as the protrusion height. Another equivalent way to determine penetration depth is to create ridges on concentric cylinder geometries and introduce continuous flow by rotating the ridged geometry with respect to the fluid. We use the second method to generate different flow patterns. Due to the introduction of the ridged geometries, the effective gap size ${R_\textnormal{gap}}$ changes as a function of the ridge spacing $S$ and the ridge depth $D$, which is denoted as
\begin{equation}
\label{Rgap_Hridges_exp}
R_\textnormal{gap}(S,D)= R_o - (R_i - \delta)
\end{equation}
where $R_i$ is assumed to be identical to the radius of the inner cylinder without the cutouts (as in Eq.~(\ref{shearrate_exp})). Due to the penetration of the flow in between the ridges, the effective gap size increases. 
The ridged geometry can thus be considered similar to the standard solid CC geometry with an effective inner radius $R_{i,\textnormal{eff}}=R_i - \delta$.  

For every geometry with a particular ridge spacing $S$ and ridge depth $D$, the penetration depth $\delta$ can be measured from standard rheology experiments.  
Rewriting the torque-rotation rate conversion equation (Eq.~(\ref{torqueeqn})), we get
\begin{equation}
M = \frac{4 \pi \eta \omega_i L (R_o R_{i,\textnormal{eff}})^2}{R_o ^2 - R_{i,\textnormal{eff}}  ^2} \hspace{0.5cm} \text{where} \hspace{0.5cm} R_{i,\textnormal{eff}} = R_i - \delta
\label{torqueeqn_ridged}
\end{equation}
In this equation, $R_i$, $R_o$, $L$ are known from the geometry and $\Omega_i$, $M$ were measured from the rheology experiments. $\eta$ was determined by fitting the flow curves in a non-ridged (solid) CC geometry with $\sigma = \eta\dot{\gamma}$. The penetration depth for every rotation rate can be obtained by solving Eq.~(\ref{torqueeqn_ridged}). For Newtonian fluids, the penetration depth does not depend on the rotation rate $\Omega_i$, therefore, we averaged the penetration depths corresponding to ten higher rotation rates to give a single penetration depth value for a geometry with a specific ridge spacing and ridge depth.

From the simulations in the flat plate limit, the penetration depth was calculated from the average shear stress on the outer wall using $\sigma=\eta\dot{\gamma}_\textnormal{eff}=\eta\Delta v/(R_\textnormal{gap,0}+\delta)$.

\section{\label{RnD}Penetration of the flow profiles between the ridges}

\begin{figure}[htbp!]
\centering
\includegraphics[width=1.0\linewidth]{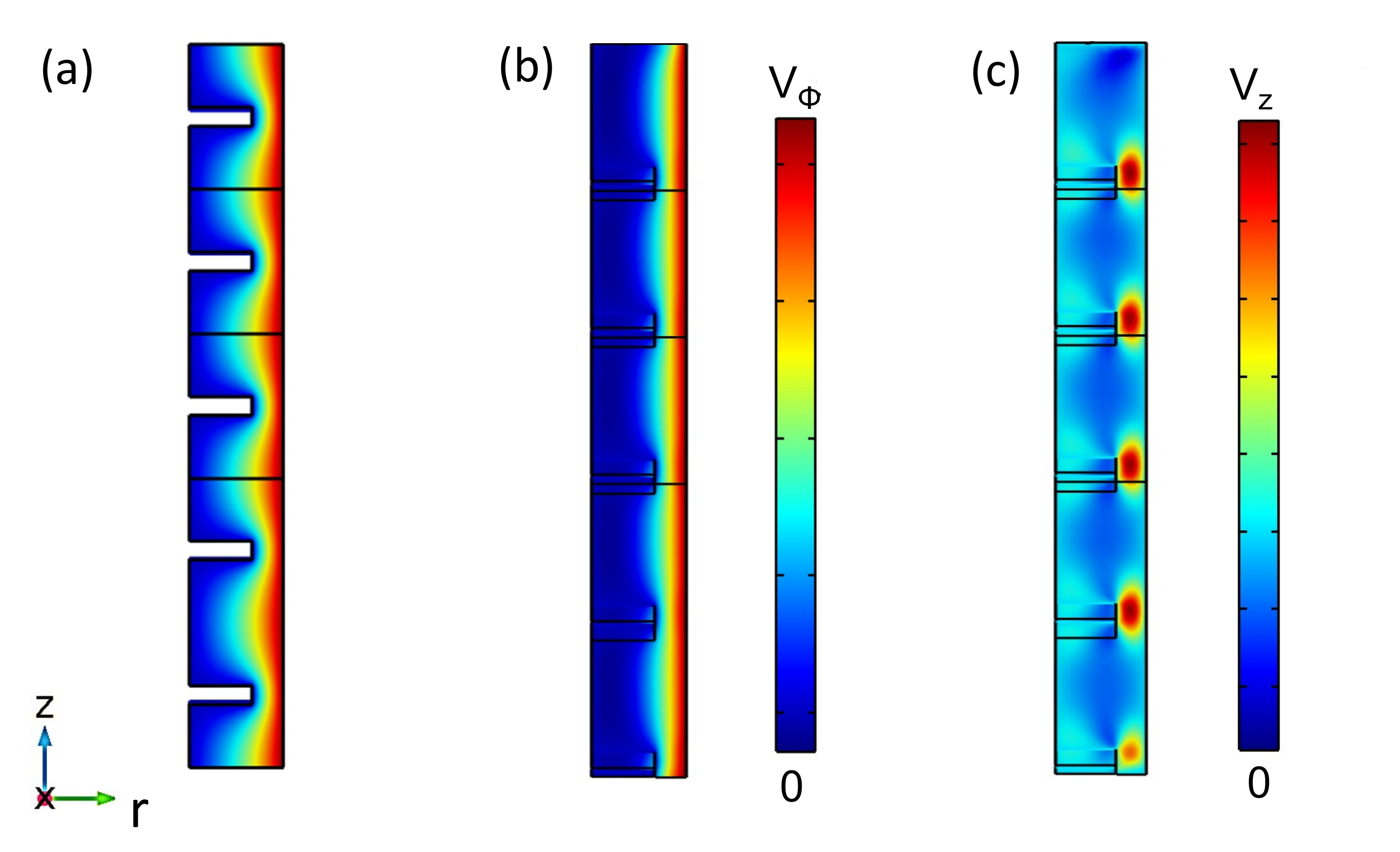}
\caption {Simulated flow profiles of a Newtonian fluid in different ridged geometries: (a) horizontal (b)-(c) angled ridges. Colorbar shows velocity in the respective directions. Here the azimuthal and axial velocities are of the order of $\sim$ 10$^{-3}$ and 10$^{-4}$ in m s$^{-1}$, respectively. For these calculations, the curved geometries were approximated as flat (infinite radius) parallel plate geometries.}
\label{simulated_flow_profiles_all}
\end{figure}

We investigated the effect of different ridged geometries on the flow behaviour of a Newtonian fluid. We numerically predicted the flow profiles for the three cases of ridged geometries: horizontal, vertical and angled ridges using finite element simulations. In the horizontal ridge case, the imposed flow direction is parallel to the ridges and the flow is purely in the azimuthal directon while in the vertical ridge case, the azimuthal flow is perturbed by the ridges, and the flow includes a radial velocity component. Finally, in the angled ridge case, the ridges exert an axial force on the fluid, therefore, there is also an axial velocity component (Fig.~\ref{simulated_flow_profiles_all}(c)). 

From the azimuthal velocity profiles (Fig.~\ref{simulated_flow_profiles_all}(a,b)), we observed that the fluid flow profile penetrates into the gap between the ridges up to a certain extent. The extent of this penetration is denoted by a parameter that we call the penetration depth ($\delta$). This flow profile penetration will influence the torque measured on the rheometer, thus, by measuring the torque, we can quantify the penetration depth as described in Subsection \ref{delta_calcmethod}. We then characterised the penetration depth for different ridge spacing, ridge depth and ridge angles.

We first evaluated the penetration depth for the two limiting cases of horizontal and vertical ridges, i.e. with ridges that are at 90$^\circ$ and 0$^\circ$ angles with respect to the vertical axis, respectively.

\subsection{Limiting case I: Horizontal ridges}

\begin{figure}[htbp!]
\centering
\includegraphics[width=1.0\linewidth]{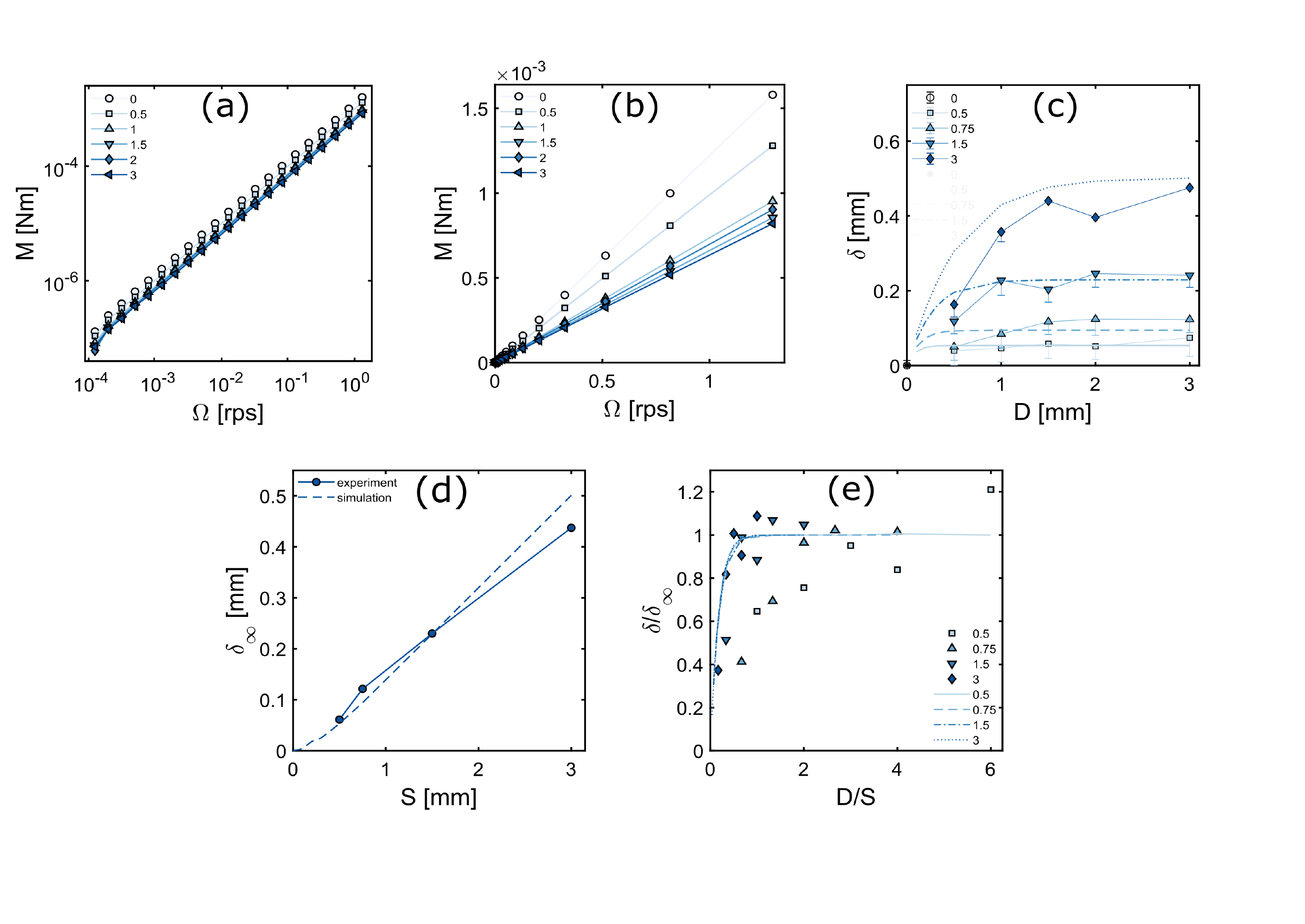}
\caption {Experimental flow curves in (a) double logarithmic scale and (b) linear scale of castor oil measured in different horizontal ridged geometries. For all geometries, $S$ = 3 mm and $D$ is indicated in the legend. (c) Variation of $\delta$ in a horizontal ridged geometry with ridge depth for different ridge spacing (legends) Experiments: symbols+lines, simulations: lines. (d) Plateau value of $\delta$ as a function of ridge spacing. A linear fit through the origin give a of slope of 0.18. (e) Master curve showing the collapse of the curves onto one exponential function when plotted as normalised $\delta$ versus normalised ridge depth. Experiments: symbols, simulations: lines. }
\label{hori_experiments+numerics}
\end{figure}

We measured the flow curves from the rotational tests on multiple horizontal ridged geometries. On a log-log scale, the torque  $M$ increased linearly with the rotation rate $\Omega$ for over four orders of magnitude in rotation rate (Fig.~\ref{hori_experiments+numerics}(a)), indicating that the presence of ridges did not change the Newtonian shear rate dependence of the flow. To investigate the variations due to the different roughness parameters, we looked more closely on a linear scale (Fig.~\ref{hori_experiments+numerics}(b)) and found that the measured torque decreases with increasing ridge depths $D$ for a given ridge spacing $S$. This is due to the influence of the flow profile penetration on the measured torque, i.e., the penetration depth ($\delta$) increases with ridge depth, leading to an increase in the effective gap size and a decrease in the measured torque. It should be noted that in all these experiments, the rotation rates  were low enough to avoid  Taylor vortices.  

Fig. ~\ref{hori_experiments+numerics}(c) shows the penetration depth, obtained from the measured torques using Eq.~(\ref{torqueeqn_ridged}), as a function of $D$ for various ridge spacings $S$. We see that $\delta$ first increases with $D$, but then levels off to a plateau that depends on the value of $S$. Hence, the flow can penetrate between the ridges only up to a certain maximum depth, $\delta_\infty$, which increases approximately linearly with the ridge spacing $S$ (Fig.~\ref{hori_experiments+numerics}(d)).

To complement the experiments and to gain more insight into the flow profile penetration, finite element simulations were performed. The geometries were recreated as described in Subsection \ref{simulations}, and the curvature of the geometry was neglected. For this geometry, the only non-zero velocity component is in the azimuthal ($\phi$) direction, while all gradients in the $\phi$-direction are zero. The pressure was also constant in this case. 

The simulated penetration depths as a function of different ridge depths and ridge spacings is plotted in Fig.~\ref{hori_experiments+numerics}(c) as dashed lines. The experimental and simulation results match quite well, especially at larger $D$. Similar to the experiments, the $\delta$ values plateau after a certain ridge depth; as in the experiments, the plateau value $\delta_\infty$ is found to increase linearly with $S$, see Fig.~\ref{hori_experiments+numerics}(d).

For a solid cylinder case ($S=0$), $\delta_{\infty}$ must be zero as the fluid cannot penetrate into a solid cylinder, and therefore, we would normally expect $\delta_{\infty}$ to scale linearly with the ridge spacing and the linear plot (dashed line in Fig.~\ref{hori_experiments+numerics}(d)) to extrapolate to the origin. However, this is not observed and $\delta_{\infty}$ seems to deviate from the linear dependence on $S$ at small ridge spacings below 0.5 mm. This is because in this small $S$ limit, the ridge spacing becomes comparable to the ridge width, which also starts to play a role in determining the penetration depth. Strikingly, the normalised penetration depth $\delta/\delta_\infty$ for different ratios of $D/S$ collapses onto a master curve (Fig.~\ref{hori_experiments+numerics}(e)). This master curve can be described by a simple exponential function:  
\begin{equation}
\label{Hori_mastercurve_eqn}
\frac{\delta}{\delta_{\infty}}=1- e^{kD/S} \hspace{0.5cm} \text{with} \hspace{0.5cm} k \approx-6.65
\end{equation}
The initial part of this master curve, corresponding to small $D$ or large $S$, can be linearized to give $\delta/\delta_\infty\approx kD/S$. It is expected that $\delta\rightarrow D$ in this regime, so that we expect $k=1/k_2$ with $k_2$ the slope of the curve of $\delta_\infty$ versus $S$. We find $1/k_2\approx 5.7$ from Fig.~\ref{hori_experiments+numerics}(d), which is indeed close to (but slightly below) the value found for $k$.
This can also be understood as follows: a static boundary layer is created due to the presence of the small ridges. The thickness of this static boundary layer is $D-\delta$ and it goes to zero as the spacing between the ridges becomes very large. The prefactor $k$ takes into account the other geometric effects, most likely the effect from the other length scale, i.e., the ridge width $w$. At a constant ridge spacing $S$, the thicker the ridges, the lower will be the flow penetration, hence, $\delta$ will be smaller. We expect that as the ridge width increases, $\delta$ will be small and hence, $\delta_\infty$ will also drop. If the normalised penetration depth $\delta/\delta_\infty$ decreases, then $k$ increases. But if both $\delta$ and $\delta_\infty$  decrease by the same factor, then $\delta/\delta_\infty$ will not change, hence, $k$ will not change. Therefore, $k$ can be written as a function of normalized $w$ as: $k = f(w/S)$.

For smaller $D$, experimental results deviate from that of simulations and this is a consistent deviation for all ridge spacings. There are some possible sources of error. It could be that the small inaccuracies during the fabrication of the geometries have a bigger influence at smaller cutouts. Moreover, the extraction of penetration depths from experiments is affected by several sources of error caused while measuring the actual inner and outer radii and the length of the cylindrical geometry and also due to end effects from the top and bottom conical parts of the geometry. For every geometry, the actual dimensions of the 3D printed geometries were measured. The contribution of end effects to the torque was calibrated on a solid cylindrical geometry. The torques for the ridged geometries were corrected for these end effects. Additionally, boundary conditions like wall slip and end effects are uncontrolled in the experiments especially for large $S$ and small $D$, which may contribute to small errors. Another possibility is that small air bubbles may be trapped between the ridges, especially for small ridge spacings. Due to these reasons, we took the plateau value of delta in experimental results as the average of the last three $\delta$ values. By linking the experiments and the simulations, it is possible to determine how changing the $D$ and $S$ of the ridges changes the penetration depth ($\delta$). 

\subsection{Limiting case II: Vertical ridges}

The second limiting case for patterned walls is a cylindrical geometry with vertical ridges. Here, we show only the numerical results in the limit where the curvature can be neglected (i.e. for large cylinder radii), see Subsection \ref{simulations}. In this case, the fluid velocity has components both in the $\phi$ and the $r$-directions, and gradients in the $z$-direction are zero. 

\begin{figure}[htbp!]
\centering
\includegraphics[width=1.0\linewidth]{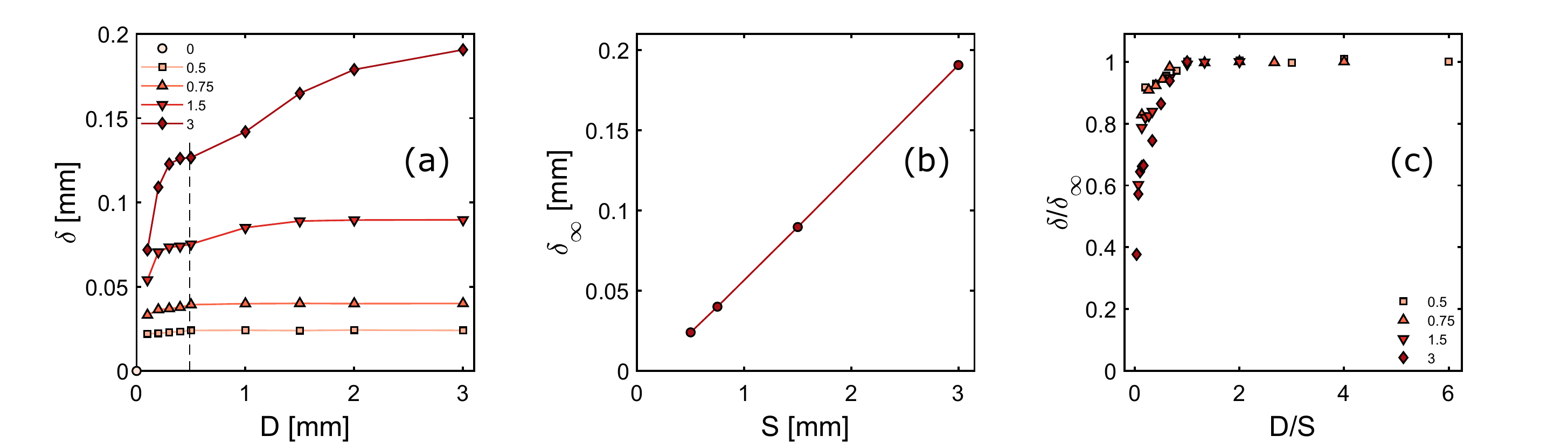}
\caption {(a) Variation of simulated $\delta$ in a vertical ridged geometry (considered as flat plate) with ridge depth and different ridge spacing (legends). Here we see a first plateau at the dashed line, followed by a second plateau at large ridge depths. (b) Plateau value of $\delta$ as a function of ridge spacing (slope = 0.067). (c) Normalised $\delta$ as a function of normalised ridge depth.}
\label{vert_flat_numerics}
\end{figure}

The penetration depth for vertical ridges with different ridge depths and ridge spacings was calculated in the same way as before and plotted in Fig.~\ref{vert_flat_numerics}(a).  As for the horizontal ridges, the flow profiles penetrate more deeply in between the ridges with increasing ridge depth and spacing, therefore, $\delta$ increases. The penetration depth values are smaller than for the horizontal ridge case because the main flow direction is perpendicular to the ridges while for the horizontal ridges, the flow is along the ridges. Also, contrary to the horizontal ridge case, we see two (quasi-)plateau values of $\delta$ as a function of $D$: the first plateau value is at around $D$ = 0.5 mm, where the ridge depth is close to the width of the ridge and most likely, the plateau value is caused by the effect of the ridge width. The second plateau value is at large ridge depths similar to what was observed in case of horizontal ridges. We consider the second plateau value and denote this as $\delta_{\infty}$.

As for the horizontal ridges, $\delta_{\infty}$ increases linearly with the ridge spacing $S$, but the slope is about 2.5 times smaller  (Fig.~\ref{vert_flat_numerics}(b)). As observed in the previous case, the linear plot does not extrapolate exactly to the origin and there seems to be non-linear dependence at small ridge spacings below 0.5 mm.

The normalised penetration depth $\delta/\delta_\infty$ for different ratios of $D/S$ again collapses quite well onto a master curve (Fig.~\ref{vert_flat_numerics}(c)). However, in this case, a single exponential fit does not work so well compared to what was seen in the horizontal case and there seems to be a two step function fit. This is mostly due to the presence of two plateaus in the $\delta$ vs D variation (Fig.~\ref{vert_flat_numerics}(a)). For $S\gg D$ i.e. for small values of $D/S$, we again expect $\delta\rightarrow D$; the plot is linear in this region and the slope of this plot is inversely related to the slope of the variation of $\delta_{\infty}$ with $S$. 

Lastly, the relation between $\delta_{\infty}$ and $S$ might depend on the width of the ridges $w$ and on the gap size ($R_o-R_i$). There are a few length scales to explore, which may give different regimes (for example: two plateau values in $\delta$ for vertical ridges). The current plots shown are for $S \gg w$. This is a very general rescaling that to our knowledge has not been reported in literature, and it is helpful to get a better understanding of the angled ridge case, as discussed below.

\subsection{Angled ridge case}

\begin{figure}[htbp!]
\centering
\includegraphics[width=1.0\linewidth]{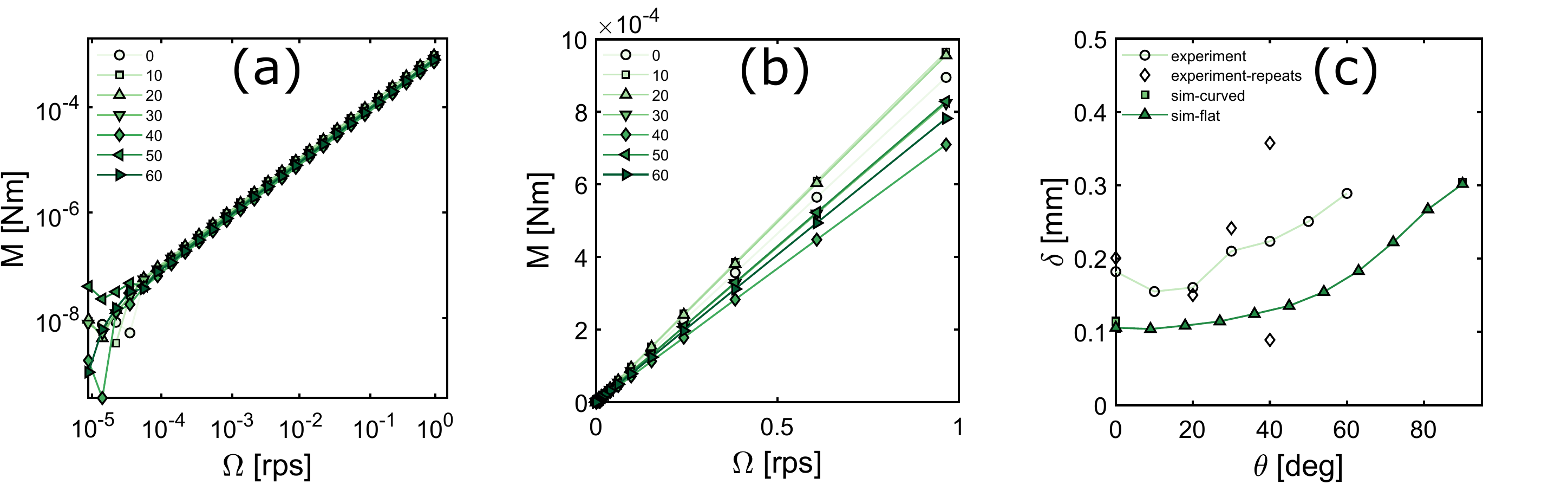}
\caption {Experimental flow curves in double logarithmic scale (a) and linear scale (b) of castor oil measured in angled ridged geometries with ridges at various $\theta$ with respect to the vertical axis. For all geometries, ridge spacing ($S$) = 2 mm  and ridge depth ($D$) = 1 mm. (c) Variation of $\delta$  as a function of the ridge angle. The simulation results are for the two limiting cases with the same ridge spacing ($S$) = 2 mm and ridge depth ($D$) = 1 mm.}
\label{ang_experiments+numerics}
\end{figure}

We can now look at the influence of geometries with ridges at finite angles on the flow behaviour of a Newtonian fluid. For different angles, the torque required to sustain a certain $\Omega$ again shows a linear dependence on the rotation rate even at higher rotation rates, indicating that the angled ridges do not affect the Newtonian shear rate dependence of the flow (Fig.~\ref{ang_experiments+numerics}(a)). All the angled ridges have the same ridge spacing and ridge depth. However, the slope of the torque versus rotation rate  changes as a function of ridge angle (Fig.~\ref{ang_experiments+numerics}(b)), indicating that the extent of flow profile penetration into the gap between the ridges changes with the angle.  To examine the effect of different angles, we calculated the penetration depths from the $M$ vs $\Omega$ relationship as previously done for the limiting cases.
 
Interestingly, $\delta$ shows a monotonic dependence on the ridge angle (Fig.~\ref{ang_experiments+numerics}(c)) and increases as $\theta$ increases, with the largest $\delta$ for the horizontal ridges ($\theta=90^\circ$). There are a few outliers at smaller ridge angles. One of the plausible reasons is $\delta$ is derived from the assumption that there is a shear stress only in the azimuthal direction. In reality, there is a normal force (and hence, a shear stress in the axial direction) and it is not clear whether the normal force is completely decoupled from the azimuthal shear stress. 

We again compare our experimental results with numerical results. In this case, the velocity has non-zero components in the $\phi$, $r$, and $z$-directions, and also the gradients in all three directions are non-zero, so, a full 3D-calculation is necessary. As we restrict ourselves to a flat geometry (corresponding to large cylinder radii), we can make use of symmetries to limit the computational domain. In particular, we model a domain with five ridges and impose a zero-outflow condition at the bottom and top of the domain, and periodic boundary conditons along the azimuthal directions. To reduce the influence of end effects, we exclude the top and bottom regions when calculating the average shear stress on the wall.

As shown in Fig.~\ref{ang_experiments+numerics}(c)), the simulated $\delta$ values for angled ridges in the flat plate regime follow a similar trend as the experimental values, with a $\delta$ that increases monotonically when the angle increases from 0$^\circ$ (vertical) to 90$^\circ$ (horizontal). However, the experimental values are systematically above the simulated ones, with an offset of $\sim$ 0.1 mm. It is not clear what determines this offset; potential factors include partial wall slip (which would reduce the measured torque), curvature effects, or end effects.

%

\section{\label{Conclusions}Conclusions}

We have introduced textured surfaces in the form of ridged geometries that can be readily fabricated using 3D printing technique. This way of patterning gives more freedom to explore an entire range of wall roughness features that can be easily characterised. These ridged geometries introduce an anisotropic wall roughness and we have shown that this has an influence on the flow behaviour in case of simple Newtonian fluids even at low Reynolds number. Due to the presence of ridges, the flow profiles penetrate in between the ridges and the extent of this penetration was defined as penetration depth ($\delta$). We have measured how this penetration depth depends on the various roughness parameters ($D$, $S$, $\theta$) from standard rheological measurements and we found a general scaling onto an exponential function. The characteristic flow behaviour of the fluid did not change due to the ridges but their magnitudes were affected. We also performed complementary simulation studies to support our experiments. 

\section*{Acknowledgments}
The authors thank the industrial partners and collaborators
for their useful comments. We thank Bob Mulder for introducing 3D
printing in rheology to us. 
This work is a part of the Industrial Partnership Programme Controlling Multiphase Flow (Project No. WP-30-01) that is carried out under an agreement between Shell, Unilever Research and Development B.V., Evodos and the Netherlands Organisation for Scientific Research (NWO). This project is co-funded by the Dutch Research Council (NWO) and TKI-E$\&$I with the supplementary grant `TKI-Toeslag' for Top Consortia for Knowledge and Innovation (TKI’s) of the Ministry of Economic Affairs and Climate Policy. This work is within the framework of the Institute of Sustainable Process Technology.


\bibliography{walleffects_Newtonian}

\end{document}